\documentclass[prd, aps, superscriptaddress, preprintnumbers, twocolumn, floatfix, nofootinbib]{revtex4}
\pdfoutput=1

\usepackage{amsfonts}
\usepackage{amsmath}
\usepackage{amssymb}
\usepackage{bm}
\usepackage{dcolumn}
\usepackage{graphicx}   
\usepackage[latin1]{inputenc}
\usepackage{latexsym}
\usepackage{rotating}
\usepackage{hyperref}
\usepackage{graphicx}
\usepackage{color}

\newcommand\bg{\begin{equation}}
\newcommand\ba{\begin{eqnarray}}
\newcommand\nd{\end{equation}}
\newcommand\ea{\end{eqnarray}}

\begin{document}

\title{Four-dimensional de Sitter space is a Glauber-Sudarshan state in string theory}

\author{Suddhasattwa Brahma}
\email{suddhasattwa.brahma@gmail.com}
\affiliation{Department of Physics, McGill University, Montr\'{e}al, QC, H3A 2T8, Canada}

\author{Keshav Dasgupta}
\email{keshav@hep.physics.mcgill.ca}
\affiliation{Department of Physics, McGill University, Montr\'{e}al, QC, H3A 2T8, Canada}

\author{Radu Tatar}
\email{Radu.Tatar@Liverpool.ac.uk}
\affiliation{Department of Mathematical Sciences,
University of Liverpool,  Liverpool, L69 7ZL, United Kingdom}

\date{\today}

\begin{abstract}

We show that four-dimensional de Sitter space is a Glauber-Sudarshan state, \textit{i.e.} a coherent state, over a supersymmetric solitonic background in full string theory. We argue that such a state is only realized in the presence of temporally varying degrees of freedom and including quantum corrections, with supersymmetry being broken spontaneously. On the other hand, fluctuations over the resulting de Sitter space is governed by the Agarwal-Tara state, which is a graviton (and flux)-added coherent state. Once de Sitter space is realized as a coherent state, and \textit{not} as a vacuum, its ability to remain out of the swampland as well as issues regarding its (meta)stability, vacuum energy, and finite entropy appear to have clear resolutions. 

\end{abstract}

\pacs{98.80.Cq}
\maketitle


The search for four-dimensional de Sitter (dS) has led to claims varying from having numerous solutions \cite{kklt,kklt1,kklt2,kklt3} to having none \cite{vafa,vafa2,vafa3}. The standard picture of a landscape of vacua \cite{Bousso:2000xa}, along with an anthropic principle to explain our own universe \cite{Weinberg:2000yb}, runs into problems not because of any no-go theorem \cite{nogo,nogo1,nogo2}, or technical hindrances \cite{Danielsson:2018ztv, Sethi:2017phn}, but due to obstructions against dS spacetimes ingrained in some fundamental aspects of quantum gravity \cite{TCC, TCC2, TCC3, Rudelius:2019cfh, EISwamp, Dvali:2018fqu}. These include trans-Planckian issues \cite{martin} challenging the very notion of a well-defined, unitary Wilsonian effective action for accelerating backgrounds \cite{TCC, TCC2, TCC3}, or that of the instabilities of the vacuum state for dS leading to a \textit{quantum swampland} \cite{Danielsson:2018qpa}. It was recently shown \cite{desitter2,desitter21,desitter22,desitter23,maxim} that the hurdles related to an effective field theory description of dS, as summarized in \cite{Palti:2019pca}, can, in principle, be overcome once \textit{time-dependent} degrees are switched on. In this work, we show that other fundamental obstructions are also avoided if four-dimensional dS space itself is regarded as a {\it state} over a Minkowski spacetime. Although similar setups have been considered before \cite{dvali,dvali1}, what is new here is an explicit construction of such a configuration realized in full string theory\footnote{A more detailed version, with proofs and computations, appears in \cite{toappear}.}. Remarkably, the resulting dS space hints at natural resolutions to problems related to its stability, vacuum energy and entropy.

We shall focus on dS in the {\it flat} slicing, with a metric of the form:
\bg\label{flatu}
{\mathrm d} s^2 = {1\over \Lambda \vert t\vert^2}\left(-{\mathrm d}t^2 + {\mathrm d}x_1^2 + {\mathrm d}x_2^2 + {\mathrm d}x_3^2\right), 
\nd
where $\Lambda$ is the cosmological constant and the temporal coordinate $t$ has a range $-\infty \le t \le 0$, with the late time regime given by $t\rightarrow 0$. We choose this specific realization of dS \eqref{flatu} only for computational efficiency and other realizations, specifically of Kasner-type, have been considered earlier \cite{desitter2}. The question is how to realize a metric like \eqref{flatu} in, say, type IIB string theory? Our first conjecture is the following metric configuration:
\bg\label{iibmet}
{\mathrm d}s^2 = {1\over \Lambda {\rm H}^2(y) \vert t\vert^2}\left(-{\mathrm d}t^2 + {\mathrm d}x_i^2\right) + {\rm H}^2(y) g_{\rm MN}(y) {\mathrm d}y^{\rm M} {\mathrm d}y^{\rm N},
\nd
where ${\rm H}(y)$ is the warp-factor that only depends on the coordinate of the internal six-dimensional manifold whose {\it unwarped} metric is given by $g_{\rm MN}(y)$. Crucially, note that the internal space is time-independent invoking the question: What kind of fluxes are required to support a configuration like \eqref{iibmet} in IIB string theory? To answer this, we uplift the configuration to M-theory. This uplift is only for computational advantage, and has no deeper implications as the degrees of freedom remain unchanged. The uplifted metric becomes:
\bg\label{mup}
{\mathrm d}s^2 = g_s^{-8/3} \eta_{\mu\nu} {\mathrm d}x^\mu {\mathrm d}x^\nu + g_s^{-2/3} {\rm H}^2 g_{\rm MN} {\mathrm d}y^{\rm M} {\mathrm d}y^{\rm N} + g_s^{4/3} 
\vert {\mathrm d}z \vert^2, 
\nd
where $g_s \equiv \sqrt{\Lambda} \vert t \vert {\rm H}(y)$ is the IIA string coupling which is now a function of both $y^{\rm M}$ and $t$, implying that in M-theory the internal manifold becomes time-{\it dependent} and $g_s \rightarrow 0$ denotes {\it late time}. Additionally, since we require ${\left(g_s/ {\rm H}\right)} < 1$, our analysis would only make sense in the interval: 
\bg\label{interval}
-{1\over \sqrt{\Lambda}} < t < 0. 
\nd
Beyond this limit, we lose all quantitative control on the dynamics. Incidentally, this means that not only is our solution trustworthy only below the \textit{quantum break-time} of dS \cite{Dvali:2018fqu} and the TCC limit \cite{TCC}, but it also provides a simple resolution of the coincidence problem of dark energy by limiting the age of our universe by the current Hubble time \cite{Velten:2014nra}, and avoiding issues of Boltzmann brains.

The dynamics is controlled by metric and fluxes because, to support such a manifold, we need G-fluxes on an eleven-dimensional space whose topology is:
\bg\label{topol}
{\cal M}_{11} = \mathbb{R}^{2, 1} \times {\cal M}_8 \equiv \mathbb{R}^{2, 1} \times {\cal M}_4 \times {\cal M}_2
\times {\mathbb{T}^2\over {\cal G}}, 
\nd
where $(\mu, \nu)$ parametrize $\mathbb{R}^{2, 1}$, $({\rm M, N})$ parametrize the six-dimensional base with unwarped metric $g_{\rm MN}(y)$ and $z^a$ parametrizes the toroidal fibre modded out by the isometry group ${\cal G}$. Let ${\bf G}_{\rm MNPQ}$ and ${\bf G}_{{\rm MN}ab}$, where $({\rm M, N}) \in {\cal M}_6$ and $(a, b) \in {\mathbb{T}^2\over {\cal G}}$, denote the $4$-fluxes switched on in the internal space. Fluxes on a {\it compact} space cannot be arbitrary: they have to satisfy Gauss' law (or alternatively, cancel anomalies), solve the EOMs, and be {\it quantized}. The quantization condition is \cite{wittenflux}:
\bg\label{wittenq}
\left[{{\bf G}_4 \over 2\pi}\right] - {p_1(y)\over 4} \in \mathbb{H}^4\left(y, \mathbb{Z}\right), 
\nd
where ${\bf G}_4$ is a four-{\it form} G-flux component, $p_1(y)$ is the first Pontryagin class and  $\mathbb{H}^4(y, \mathbb{Z})$ is the fourth cohomology class. In a time-dependent internal space like \eqref{mup}, this quantization condition cannot be satisfied with time-{\it independent} G-flux components. Additionally EOMs will require fluxes to become time-{\it dependent} raising, in turn, serious questions about how could these fluxes be quantized, and how could they satisfy Gauss' law. On the dual IIB side, this means that the three and the five-form fluxes should become time-{\it dependent}, leading to similar questions. The axio-dilaton however remains time-independent because we are in the constant-coupling scenario of F-theory \cite{DM, desitter2}. In the absence of quantum corrections, as shown in \cite{desitter2}, all the questions raised above cannot be answered and the system is inconsistent, stemming from the loss of $g_s$ and ${\rm M}_p$ hierarchies, leading to a breakdown of effective field theory as a manifestation of the swampland conjectures \cite{Palti:2019pca}. More importantly, a metric like \eqref{mup} is still a bit too constrained to be a solution, and what really works is a metric of the form:
\begin{eqnarray}\label{mup2}  
{\mathrm d}s^2 &=& g_s^{-8/3} \eta_{\mu\nu} {\mathrm d}x^\mu {\mathrm d}x^\nu + g_s^{-2/3} {\rm H}^2\Big({\rm F}_1(t) g_{\alpha\beta} {\mathrm d}y^\alpha {\mathrm d}y^\beta \nonumber\\
 && \;\;\;\;+ {\rm F}_2(t) g_{mn} {\mathrm d}y^{m} {\mathrm d}y^{n}\Big) + g_s^{4/3} 
\vert {\mathrm d}z \vert^2,
\end{eqnarray}
where $(\alpha, \beta) \in {\cal M}_2$ and $(m, n) \in {\cal M}_4$ of \eqref{topol}. The additional time-dependences governs how the six-dimensional manifold ${\cal M}_6 = {\cal M}_4 \times {\cal M}_2$ changes with respect to time. On the IIB side this converts \eqref{iibmet} to the following:
\ba\label{iibmet2}
{\mathrm d}s^2 &= & {1 \over \Lambda {\rm H}^2(y) \vert t\vert^2}\left(-{\mathrm d}t^2 + {\mathrm d}x_1^2 + {\mathrm d}x_2^2 + {\mathrm d}x_3^2\right) \\
&+& {\rm H}^2(y)\Big({\rm F}_1(t)  g_{\alpha\beta}(y) {\mathrm d}y^\alpha {\mathrm d}y^\beta + {\rm F}_2(t) 
g_{mn}(y) {\mathrm d}y^m {\mathrm d}y^n\Big), \nonumber
\ea
making the internal space time-dependent. To preserve four-dimensional Newton's constant, we additionally require ${\rm F}_1 {\rm F}^2_2 = 1$ and both ${\rm F}_i(t) \to 1$ as $g_s \to 0$. Imposing all these conditions, the system becomes tightly constrained but does have a solution, answering all the questions that we raised above, as shown in \cite{desitter2}. However, even after we get an effective potential which supports dS space in string theory, there still remains questions whether the radiative corrections, calculated for some vacuum state, leads to instabilities \cite{Danielsson:2018qpa} or if the obstructions regarding trans-Planckian issues \cite{TCC} lingers on. To answer these, we attempt to see if a metric like \eqref{iibmet2} or \eqref{mup2} can be realized as a {\it state} instead of a {\it vacuum}.

The state that we have in mind is the Glauber-Sudarshan state \cite{glauber, glauber1}, commonly called a coherent state, because such a state is closest to a {\it classical} configuration which can be realized in a quantum theory. Our aim would be to realize \eqref{mup2} as a Glauber-Sudarshan state over a supersymmetric solitonic vacuum. (Then, by dualization, \eqref{iibmet2} can also be realized as such a state.) The supersymmetric vacuum would be our warped Minkowski space. As we shall see there are numerous advantages from such a realization, one of them being -- because of our choice of supersymmetric vacuum -- the cancellation of the zero point energies and the subsequent spontaneous breaking of supersymmetry by the state. In fact, supersymmetry will be broken due to the presence of non self-dual G-fluxes over the eight manifold ${\cal M}_8$. Another advantage is that Wilsonian analysis {\it can} be performed because the modes in our theory are the ones over the solitonic background and not the ones  over a dS background with time-dependent frequencies. Consequently, issues such as trans-Planckian problems \cite{martin} would cease to be of any concern. The supersymmetric solitonic background that we have in mind is of the form:
\bg\label{solikaton}
{\rm d}s^2 = {1\over h^{2/3}(y)}\left(-{\rm d}t^2 + {\rm d}x^2_i\right) + h^{1/3}(y) g^{(0)}_{\rm MN} {\rm d}y^{\rm M} {\rm d}y^{\rm N}, 
\nd
where $h(y)$ is the warp-factor, and can be supported by self-dual G-fluxes of the form ${\bf G}^{(0)}_{\rm MNPQ}(y)$, as well as ${\bf G}^{(0)}_{0ij{\rm M}}$, where $({\rm M, N}) \in {\cal M}_8$. Such a background has been studied in detail in \cite{becker}. If we study {\it fluctuations} over this background, they are classified by modes which may be easily determined. These modes will typically have non-trivial spatial behavior, whose dynamics will be governed by a Schr\"odinger equation over a non-trivial potential, but their temporal behavior would be simple ($\sim e^{i \omega_{\bf k} t}$). This will help us to avoid any trans-Planckian issues, but new subtleties lie in the construction of the Glauber-Sudarshan state itself. In the original work of \cite{glauber}, the state was created by shifting the {\it free} vacuum (or the harmonic vacuum) by a displacement operator. One of the main issue with such a construction is that there is {\it no} free vacuum in a highly interacting theory like M-theory! Secondly, M-theory has metric as well as G-flux components, so a Glauber-Sudarshan state would be more complicated to account for fluctuations of all these components. Thirdly, even if we manage to construct such a state, how do we know that such a state survives the set of quantum corrections coming from perturbative, non-perturbative, non-local and topological interactions?

Let us start by answering the first question, related to the construction of the Glauber-Sudarshan state. Since there is no free, or harmonic, vacuum once interactions are switched on, we only have the {\it interacting} vacuum $\vert \Omega\rangle$ to build our state from. We shall shift the interacting vacuum by a displacement operator and ask if this creates a state resembling the Glauber-Sudarshan state. The state we want to construct is of the form:
\bg\label{coherbeta}
\vert \sigma\rangle \equiv \mathbb{D}(\sigma) \vert \Omega \rangle,
\nd
at a specific time $t = t_0$. $\mathbb{D}(\sigma)$ is the {\it displacement} operator and $\sigma \equiv \left(\{\alpha_g\}, \{\beta_{\rm C}\}\right)$ where $\{\alpha_g\}$ and $\{\beta_{\rm C}\}$ are the two sets of parameters associated with all the metric components and all the C-field components, respectively. The meaning of a displacement operator in an interacting theory is, however, not clear. In the free theory, a displacement operator is constructed from annihilation $a_{\bf k}$ and creation $a_{\bf k}^\dagger$ operators for a given spatial momentum ${\bf k}$, so in an interacting theory we expect analogous operators $a_{\rm eff}({\bf k})$ and $a^\dagger_{\rm eff}({\bf k})$ to replace them, with $a_{\rm eff}({\bf k})$ annihilating the interacting vacuum. Unfortunately this information is not enough to fix the form of $a_{\rm eff}({\bf k})$, which is a complicated function that mixes the free-field annihilation and the creation operators for different spatial momenta \cite{toappear}. Because of that, the displacement operator $\mathbb{D}(\sigma)$ is {\it not} a unitary operator anymore. However, there does exist one possible choice for $\mathbb{D}(\sigma)$ that not only fixes the form for $a_{\rm eff}({\bf k})$, but also reproduces the background \eqref{mup2} as expectation values of the metric operators in the state \eqref{coherbeta}. This choice works for any time $t$ and is given by:
\bg\label{coher2} 
\mathbb{D}(\sigma, t) = \mathbb{D}_0(\sigma)~{\rm exp}\left(i \int_{-T}^t d^{11}x ~{\bf H}_{\rm int}\right), 
\nd
where $T \to \infty$ in a slightly imaginary direction; and ${\bf H}_{\rm int} \equiv {\bf H}_{\rm int}(g_{\rm MN}, {\rm C}_{\rm PQR})$ is the full interacting part of the M-theory Hamiltonian. $\mathbb{D}_0(\sigma)$ is the displacement operator for the harmonic vacuum, meaning that it displaces the harmonic vacuum by $\sigma \equiv \sigma(t)$ to create the required Glauber-Sudarshan state with one minor difference: it is the non-unitary part of the usual free vacuum displacement operator. In writing \eqref{coher2} we have ignored a multiplicative constant piece that is proportional to the {\it overlap} between the interacting and the harmonic vacuum, \textit{i.e.} $\langle \Omega \vert 0 \rangle$. One can also work out the wavefunction in the configuration space for the state \eqref{coherbeta} satisfying \eqref{coher2}. For example, say, for the spacetime mode for the graviton, the wave-function of the state \eqref{coherbeta} may be expressed explicitly as:
\bg\label{wavefunction}
\Psi^{(\sigma)}_\Omega(g_{\mu\nu}, t) = {\rm exp}\left[\int_{-\infty}^{+\infty} d^{10}{\bf k} 
~{\rm log} \langle \widetilde{g}_{\mu\nu}({\bf k})\vert \Psi_{\bf k}^{(\sigma)}(t)\rangle\right], 
\nd
where $\vert \Psi_{\bf k}^{(\sigma)}(t)\rangle$ is the Glauber-Sudarshan state in the Heisenberg representation for a given spatial mode ${\bf k}$ and $\widetilde{g}_{\mu\nu}({\bf k})$ is the Fourier component of the graviton.

Indeed, this is all we need for the present purpose, as all of the background quantities simply appear by taking expectation values over the state \eqref{coherbeta} with the wave-function \eqref{wavefunction}. For example, let us again concentrate on the space-time metric. The expectation value of the metric operator may be expressed as:
\bg\label{patho}
\langle{\bf g}_{\mu\nu}\rangle_\sigma = {\int [{\cal D}g_{\mu\nu}]e^{i{\bf S}}\mathbb{D}^\dagger(\sigma) 
g_{\mu\nu} \mathbb{D}(\sigma) \over 
\int [{\cal D}g_{\mu\nu}]e^{i{\bf S}}\mathbb{D}^\dagger(\sigma)\mathbb{D}(\sigma)}
 =  {\eta_{\mu\nu} \over \left(\Lambda \vert t\vert^2 {\rm H}^2(y)\right)^{4/3}},
\nd
where ${\bf S} \equiv {\bf S}(g_{\rm MN}, {\rm C}_{\rm PQR})$ is the total M-theory action. Since $\mathbb{D}(\sigma) \equiv \mathbb{D}(\sigma(t), t)$ is non-unitary, not only does it necessitate a division by another path integral as shown in the middle equality, but also it keeps the numerator from vanishing. In fact, $\mathbb{D}(\sigma)$ does what it is expected to do: it shifts the vacuum in such a way that the one-point functions do not vanish and ensures that they have the necessary expectation values. In addition, the choice \eqref{coher2} guarantees that there are no ${\cal O}\left({g_s^a\over {\rm M}_p^b}\right)$ corrections to \eqref{patho} \cite{toappear}. The above computation relies on two essential objects: (i) the space-time wave-functions (not the configuration space wave-functions!) which come from solving a class of Schr\"odinger equations with non-trivial potentials, and (ii) the Glauber-Sudarshan wave-function \eqref{wavefunction}. Similarly, the expectation value of the G-flux components over the state \eqref{coherbeta} becomes:
\bg\label{gflux}
\langle {\bf G}_{\rm MNPQ}\rangle_\sigma = \sum_p {\cal G}^{(p)}_{\rm MNPQ}(y) 
\left({g_s \over {\rm H}}\right)^{2p/3}, 
\nd
where $({\rm M, N}) \in {\cal M}_8$ with $p \in {\mathbb{Z}\over 2}$ and $p \ge {3\over 2}$. The latter condition stems from various criteria, including flux equations of motion, Bianchi identities and subtleties with {\it localized fluxes}, as elaborated in \cite{desitter2}. We want to emphasize that the bound on $p$ tells us that there are {\it no} time-independent fluxes allowed in this set-up, and that supersymmetry is broken spontaneously because the G-flux on the internal space is no longer self-dual, \textit{i.e.}
\bg\label{noself}
\vert \langle {\bf G}_4\rangle_\sigma - \langle \ast_8 {\bf G}_4 \rangle_\sigma\vert > 0, 
\nd
where ${\bf G}_4$ is the four-form operator and $\ast_8$ is the Hodge dual with respect to the un-warped metric of the solitonic background \eqref{solikaton}. Note that the fluxes ${\bf G}_4^{(0)}$, supporting \eqref {solikaton}, are self dual and preserve supersymmetry for the solitonic vacuum and, therefore, the supersymmetry is broken spontaneously by the Glauber-Sudarshan state.

Next, we wish to understand the {\it fluctuations} over dS space in such a setting. Clearly, since dS itself is a state over the solitonic vacuum, the fluctuations should also come from a related state appearing as some deformation of the  Glauber-Sudarshan state. It turns out that the required deformation is another well-known state, called the {\it Agarwal-Tara} state \cite{agarwal}, or alternatively, as the graviton added coherent state. For us, this needs to be generalized in the same vein as the Glauber-Sudarshan state so that the Agarwal-Tara state will have to be both graviton and flux-added coherent state, which is given by:
\bg\label{tara}
\left\vert \Psi_{\bf k}^{(c_1c_2)}(t)\right\rangle = \left[c_1 + c_2 {\cal G}\left(a_{\bf k} + a^\dagger_{\bf k}; t\right)
\right] \left\vert \Psi^{(\sigma)}_{\bf k}(t)\right\rangle,
\nd
where we have used \eqref{coher2}, and denote $a_{\rm eff}({\bf k})$ and $a^\dagger_{\rm eff}({\bf k})$ by $a_{\bf k}$ and $a^\dagger_{\bf k}$  to simplify the notation, and $c_i$ are constants with $|c_2| << |c_1|$. We have further restricted to the graviton sector to avoid over-burdening the formula with complications from the flux sector. ${\cal G}(w, t)$ could be thought of as a polynomial function of $w$ with time-dependent coefficient (details appear in \cite{toappear}). Subtleties aside, what is significant for us is that the expectation of the graviton operator in such a state gives us:
\ba\label{ATs}
&&\langle {\bf g}_{\mu\nu} \rangle_{\Psi^{(c_1c_2)}} =  
{\eta_{\mu\nu} \over \left(\Lambda \vert t\vert^2 {\rm H}^2(y)\right)^{4/3}}\\
&& ~~~~ + c_2 \int d^{11} k~ h({\bf k}, k_0)~ 
\alpha_{\mu\nu}({\bf k}, k_0)~ \psi_{\bf k}({\bf x}, y, z) ~e^{ik_0 t}, \nonumber \ea
where $\alpha_{\mu\nu}$ is precisely the parameter from the set $\{\alpha_g\}$ of $\sigma \equiv \left(\{\alpha_g\}, \{\beta_{\rm C}\}\right)$ that defines the Glauber-Sudarshan state \eqref{coherbeta}. $\psi_{\bf k}({\bf x}, y, z)$ is the eleven-dimensional spatial wave-function that appears from the Schr\"odinger equation over the solitonic background alluded to earlier. We have also taken $c_1 = 1$, so only $c_2$ appears in \eqref{ATs}. $h({\bf k}, k_0)$ is the Fourier component that comes from ${\cal G}(w, t)$ function in \eqref{tara}. The crucial take-home point from \eqref{ATs} is that the temporally varying frequencies $\omega_{\bf k}(t)$ that we get from fluctuations over a dS {\it vacuum} are nothing but {\it artifacts} of the Fourier transforms over the Glauber-Sudarshan state. This is because:
\bg\label{notcc}
{\rm exp}\left(-i\omega_{\bf k}(t) t\right) \equiv \int dk_0~h({\bf k}, k_0)~e^{-ik_0 t}, 
\nd
where $\omega_{\bf k}(t)$ is in general a complex function. Such a conclusion seems to point out that there are no trans-Planckian censorship required for our construction \cite{TCC} because all modes originate secretly from fluctuations over our solitonic background \eqref{solikaton}. (We satisfy the conditions of having a underlying Lorentz-invariant spacetime, with a local vacuum, required to avoid the original trans-Planckian problem \cite{martin}.) Since our dS space,itself, is a state which results from these modes, it is no surprise that we also get fluctuations on top of dS as another consequence of them.

It is encouraging to see that all the background information is encoded in the  expectation values of the graviton and the C-field operators over the Glauber-Sudarshan and the Agarwal-Tara states. However, we haven't tackled the question of the {\it stability} of the background \eqref{mup2}. How do we know that the quantum corrections do not take us away from the configuration \eqref{mup2} and \eqref{gflux}? In fact, any small changes to the ${\rm F}_i(t)$ factors in \eqref{mup2} will switch on some time-dependence of the Newton's constant. In order to avoid this, we have to establish the stability of the solution. This also gets tied up with the equations of motion in the presence of {\it all} possible quantum corrections. 

It turns out that the background equations may be presented most succinctly as Schwinger-Dyson equations (SDEs). The SDEs \cite{dyson, dyson1, dyson2} are expressed as expectation values over the state \eqref{coherbeta} and therefore suits our construction very well since we have considered \eqref{patho}, G-fluxes \eqref{gflux} as well as the fluctuations \eqref{ATs} as expectation values. The SDEs for our case may be divided into two set. One set is easy to write down and is given by:
\bg\label{sde}
{\delta {\bf S}^{(\sigma)} \over \delta \langle g^{\rm MN}\rangle_\sigma} = {\delta {\bf S}^{(\sigma)}\over \delta \langle{\rm C}^{\rm MNP}\rangle_\sigma} = 0,
\nd
where ${\bf S}^{(\sigma)} \equiv {\bf S}^{(\sigma)}\left(\langle g_{\rm MN}\rangle_\sigma, \langle {\rm C}_{\rm PQR}\rangle_\sigma\right)$ differs from ${\bf S} = {\bf S}(g_{\rm MN}, {\rm C}_{\rm PQR})$ in \eqref{patho} by the appearance of the expectation values in the integrands themselves. The second set of equations are however more involved and they include both the Faddeev-Popov ghosts and the displacement operator \eqref{coher2} \cite{toappear}. For us \eqref{sde} will suffice, as the form of these equations precisely imply the equations of motion already studied in \cite{desitter2} and, therefore, pursues the path laid down there, as follows. Once we express the equations order by order in $\left({g_s\over {\rm H}}\right)$, the zeroth order equations precisely determine the background \eqref{mup2} with the G-fluxes as in \eqref{gflux}. Going to the higher orders in $\left({g_s\over {\rm H}}\right)$ then switches on three things: (a) higher orders, i.e $p > {3\over 2}$ in \eqref{gflux} for the G-fluxes, (b) higher order terms for the ${\rm F}_i(t)$ factors in \eqref{mup2}, and (c) higher order quantum terms discussed in \cite{desitter2}. Together, they {\it balance} each other in such a way that the zeroth order metric and G-fluxes, from \eqref{mup2} and \eqref{gflux} respectively, {\it do not receive any corrections} \cite{toappear}.

The action ${\bf{\bf S}}^{(\sigma)}$ above contains all types of quantum corrections, for instance, non-perturbative corrections of the BBS \cite{bbs} and KKLT \cite{kklt} type instantons as well as the action of the branes and surfaces, including fermionic interactions, as shown in \cite{toappear}. Indeed, this is an important point since the cosmological constant in four-dimensions, in our formalism, emerges from a delicate balancing between the fluxes and the quantum corrections but {\it without} any vacuum energies \cite{desitter2}. Furthermore, since the fluctuations are ultimately built out of the interacting vacuum of the solitonic background, perturbative quantum corrections do not lead to any instabilities as everything is expressed in terms of time-independent mode expansions \cite{toappear}, as opposed to the case of classical dS \cite{Danielsson:2018qpa}.

These fluctuations turn out to be intimately connected to the {\it entropy} of dS spacetime, an old puzzle being the origin of its finiteness within the context of full quantum gravity \cite{Bousso:2000xa}. Specifically, in our solution, a natural resolution would be in interpreting this as the \textit{entanglement entropy} between the fluctuation modes, on top of the time-independent solitonic vacuum, which give rise to the Galuber-Sudarshan state itself. The way we get a finite von Neumann entanglement entropy, due to this mode-coupling, is that we have a reduced density matrix corresponding to tracing out the modes which are super-horizon, and treating the causal patch of an inertial observer as our system. The key observation is that if the interactions could be turned off, such an entanglement entropy would consist of diverging parts alone but, in that case, we would have no dS space either as $ {\rm {\bf H}}_{\rm int}$ is crucial in the construction of \eqref{wavefunction}. It had been heuristically argued earlier that, for a coherent state description of dS, a finite number of gravitons is synonymous to the finite entropy associated with it; furthermore, it was also noted that interactions between the gravitons are important to arrive at this conclusion \cite{Dvali:2018fqu}. In our case, we sketch out the concrete reason as to how one arrives at a finite number of highly-interacting gravitons (and flux particles) for our Glauber-Sudarshan state necessary for having a finite entropy for the resulting dS space. Our solution also satisfies the expectation of the dS symmetries being emergent for a  `reasonably' short time-period, and are \textit{not} eternal, which is a necessary condition for having a finite entropy \cite{Goheer:2002vf, ArkaniHamed:2007ky}. Without going into details, note that the above procedure is technically tractable due to the fact that the entanglement entropy corresponding to coherent states coincide with that for the vacuum \cite{toappear, Benedict:1995yp}. 

To conclude, in this work, we have shown how the EOMs can be solved in full string theory to obtain a dS spacetime by avoiding the swampland, once time-dependent degrees of freedom are turned on, and including quantum corrections, by constructing the solution in terms of a Glauber-Sudarshan state. Remarkably, this state (a) solves the trans-Planckian censorship problem, (b) is stable against both perturbative and non-perturbative quantum corrections, and (c) provides a microscopic understanding of its entropy as an inherently quantum quantity. Furthermore, the order of magnitude of the time-limit before which the system becomes strongly-coupled gives an easy way out of the coincidence problem of cosmology as well as an escape from issues related to  Boltzmann brains.

\noindent {\bf Acknowledgements:} We would like to thank Robert Brandenberger for discussions. The work of KD is supported in part by the Natural Sciences and Engineering Research Council of Canada. SB is supported in part by funds from NSERC, from the Canada Research Chair program, by a McGill Space Institute fellowship and by a generous gift from John Greig.

\end{document}